\documentclass{sig-alternate}

\usepackage[hyphens]{url} 
\usepackage{graphicx}%
\usepackage{dcolumn}%
\usepackage{bm}%
\usepackage{psfrag}
\usepackage{color}
\usepackage{amsmath, amssymb, latexsym, amsfonts, epsfig, graphicx, verbatim}
\usepackage[nospace]{cite}
\usepackage{algorithmic, algorithm}
\usepackage{verbatim}
\usepackage{xspace}
\usepackage{multirow}
\usepackage{makecell}
\usepackage{subfig}
\usepackage{caption}
\newcommand{\Var}{\mathrm{Var}}
\newcommand{\eqn}[1]{Eq.(#1)}

\newcommand{\Sec}[1]{Section~\ref{#1}}
\newcommand{\Fig}[1]{Fig.~\ref{#1}}
\newcommand{\Tab}[1]{Table~\ref{#1}}
\newcommand{\est}[1]{\widehat{#1}}
\newcommand{\ie}{{\em i.e., }}
\newcommand{\eg}{{\em e.g., }}

\definecolor{heraldBlue}{rgb}{0.0,0.0,0.8}
\definecolor{heraldRed}{rgb}{0.8,0.0,0.0}
\definecolor{heraldGreen}{rgb}{0.0,0.4,0.0}
\definecolor{heraldPurple}{rgb}{0.9,0.1,0.9}

\begin{document}

\title{Estimating Subgraph Frequencies with or without\\ Attributes from Egocentrically Sampled Data}

\numberofauthors{1}
\author{ 
\alignauthor  Minas Gjoka$^{\dagger}$, Emily Smith$^{\circ}$, Carter T. Butts$^{\dagger\circ\diamond}$\\
        \affaddr{EECS Dept.$^\dagger$, Sociology Dept.$^\circ$, Statistics Dept.$^\diamond$}\\
        \affaddr{University of California, Irvine}\\
	\email{ $\{$mgjoka, emilyjs, buttsc$\}$@uci.edu}
}

\maketitle

\begin{abstract}
In this paper we show how to efficiently produce unbiased estimates of subgraph frequencies from a probability sample of egocentric networks (i.e., focal nodes, their neighbors, and the induced subgraphs of ties among their neighbors). A key feature of our proposed method that differentiates it from prior methods is the use of egocentric data.
Because of this, our method is suitable for estimation in large unknown graphs, is easily parallelizable, handles privacy sensitive network data (e.g. egonets with no neighbor labels), and supports counting of large subgraphs (e.g. maximal clique of size 205 in \Sec{sec:facebook}) by building on top of existing exact subgraph counting algorithms that may not support sampling.  It gracefully handles a variety of sampling designs such as uniform or weighted independence or random walk sampling. Our method can be used for subgraphs that are: (i)  undirected or directed; (ii)  induced or non-induced; (iii) maximal or non-maximal; and (iv) potentially annotated with attributes.  We compare our estimators on a variety of real-world graphs and sampling methods and provide suggestions for their use. Simulation shows that our method outperforms the state-of-the-art approach for relative subgraph frequencies by up to an order of magnitude for the same sample size. Finally, we apply our methodology to a rare sample of Facebook users across the social graph to estimate and interpret the clique size distribution and gender composition of cliques.
\end{abstract}

\section{Introduction}

In a large number of real-world applications it is common to represent systems, structures, or data using graphs
\eg social graphs, web graphs, or protein interaction graphs. In many cases these graphs are difficult to study, most commonly because of their massive size  and/or access limitations. As a result, there is a growing body of work \cite{Kolaczyk2009,gjoka10_walkingfb,Hardiman2009}  that uses sampling to estimate the properties of such graphs  as a step towards understanding them.  In this paper, we show how to efficiently produce unbiased estimates of the count of an (optionally maximal) subgraph or induced subgraph of a given form in a graph or digraph from a probability sample of nodes, with or 
without
nodal attribute constraints.

There has been great interest in the research community in counting either statistically over-represented (partial or induced) subgraphs \cite{holland.leinhardt:sm:1975}, sometimes called network motifs \cite{milo2002network}, or the full census of induced subgraphs of a given size, called graphlets \cite{prvzulj2007biological} (typically 3, 4, and 5-node). In the seminal work by Holland and Leinhardt \cite{holland.leinhardt:cgs:1971}, global network structures were shown to be heavily constrained by their subgraph composition; this work also established the enumeration and labeling of graph isomorphism classes to identify specific types of subgraphs, an approach that has since become standard practice in many fields \cite{wasserman.faust:bk:1994,barabasi2004network,prvzulj2007biological}.  \cite{holland.leinhardt:jasa:1981,frank.strauss:jasa:1986} introduced use of parametric statistical models for networks based on subgraph counts, an approach that is also now in wide use \cite{hunter.et.al:jcgs:2012}.  Parametric network models based on graphlets were introduced by \cite{yaveroglu.et.al:jss:2015}, who applied them to social and protein structure networks. Particular classes of subgraphs have also found applications in a number of fields. Cliques, for instance, have been studied in social networks as the foundation for clustering and cohesive subgroups \cite{ wasserman1994social}. In addition to social networks \cite{ krackhardt2002structure, jansson1997clique, bernard1980informant}, cliques have been analyzed in a diverse range of networks, including those relating to protein structure \cite{ grindley1993identification, strickland2005optimal}, image recognition \cite{stentiford2010image}, and written texts \cite{ caldeira2006network}. Open or incomplete two-paths have been employed as motifs for the study of brokerage in social and organizational systems \cite{gould.fernandez:sm:1989,spiro.et.al:sn:2013}, and $k$-stars have been used as tools for modeling degree distributions \cite{snijders.et.al:sm:2006}.  Vital for all of these applications is the ability to count subgraphs of particular types.

\begin{figure}
\centering
\includegraphics[width=0.49\textwidth]{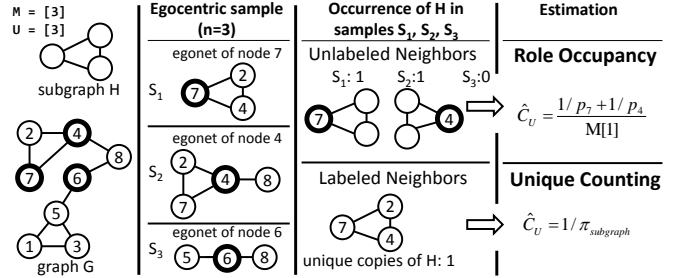}
\caption{Illustration of the egocentric approach, and the difference between labeled and unlabeled neighbors. The goal is to estimate the subgraph $H$ count in graph $G$. In this case, egos 7, 4, 6 are sampled from graph $G$. }
\label{fig:labeling}
\end{figure}

The demand for subgraph counts has spurred the development of a variety of algorithms. They can be classified in two categories: those that use exact counting \cite{chen2006nemofinder,kashani2009kavosh,grochow2007network,schreiber2005mavisto,hovcevar2014combinatorial} and those that use sampling \cite{wernicke2006fanmod, omidi2009moda, kashtan2004efficient, bhuiyan2012guise, wang2014efficiently}. 
Exact algorithms require knowledge, processing, and storage of the entire graph.  In contrast, we provide estimators for subgraph counts using only samples of network data. More importantly, our methods {\it extend support for sampling} to all exact subgraph enumeration techniques that do not support it by design. In comparison with existing sampling methods, our methods additionally support annotated attributes, and either induced or non-induced subgraphs. In 
Sec. \ref{sec:simulation}
we show that our methods outperform the state-of-the-art approach of \cite{wang2014efficiently} by up to an order of magnitude for the same sample size.

Our estimation techniques for subgraph counts employ an egocentric approach \cite{wasserman1994social}, illustrated in \Fig{fig:labeling}. We first collect a probability sample of nodes (``egos'') from the target graph. In our example we sample nodes 7, 4, 6. Then, we collect the egonet of each sampled node, which consists of the neighbors of the node and the edges between these neighbors. Next, we use an existing enumeration tool to calculate the exact subgraph count in each sampled egonet. Our approach supports the absence or presence of unique neighbor labels in the egonets. \Fig{fig:labeling} shows the implications of either possibility in the partial count. Last, depending on the existence of neighbor labeling we apply the Role Occupancy or Unique Counting estimation method to combine the subgraph counts in individual egonets and estimate the subgraph count for the whole graph in an unbiased manner.

Our approach has several benefits. The first obvious benefit is that it allows us  to estimate the subgraph count for a given isomorphism class from an unknown graph, as long as we have a sampling primitive that reveals the neighbors of a selected node in that graph. 
The second benefit is that it can be used to estimate the subgraph count in a fully known massive graph. Our approach decomposes a large problem into many smaller problems that can be independently computed, hence making estimation embarrassingly parallel.
Another benefit is the ability to estimate the subgraph count in the absence of unique neighbor node labels \eg  due to privacy-sensitive network data or data collection limitations. Finally, our techniques can be employed with data collected using standard techniques in both online (\eg random walk or user ID sampling \cite{gjoka2011practical}) and offline (\eg survey instruments \cite{marsden2005recent}) settings.

In summary, we make the following contributions:
\begin{itemize}
 \item We  present two unbiased estimation methods to efficiently estimate the frequency of a subgraph from an egocentric probability sample of nodes. Our methods support subgraphs of any order so long as they are contained within an egonet, including variations such as undirected/directed, induced/non-induced, maximal/ general cases.  Our methods also allow for counting of subgraphs that are differentiated by nodal attributes. Our first method, Role Occupancy, is applicable for egocentric data whether or not the neighbors of sampled nodes can be uniquely identified across draws (\emph{unlabeled}), and our second method, Unique Counting, provides additional statistical power where neighbors are uniquely labeled.  
  \item We evaluate our methods on a variety of real-world graphs and sampling methods and provide suggestions for their use.  Unique Counting is shown to have on average smaller error than the Role Occupancy method. However that comes with additional space complexity, which can be quite significant depending on the subgraph of interest. We show that choice of sampling method affects the estimation error, with decorrelated random walks and weighted independence sampling having the smallest error. 
  \item We apply our methodology to a sample of Facebook (FB) users to estimate the clique size distribution and gender composition of cliques across the social graph. Through our analysis we discover evidence for strong heterogeneity in the makeup of cliques.

\end{itemize}

The structure of the remainder of the paper is as follows. Section~\ref{sec:related} reviews related work. Section~\ref{sec:notation} presents the definitions and basic concepts. 
Section~\ref{sec:estimation} presents our estimation methodology.  Section~\ref{sec:simulation} presents simulation results on real-life fully known graphs.  Section~\ref{sec:facebook} applies our estimators to samples collected from Facebook.  Finally, Section~\ref{sec:conclusion} concludes the paper.

\section{Related Work}
\label{sec:related}

Egocentric sampling is a widely used method for gathering network data \cite{marsden1990network,carrington2005models}. This method samples individual nodes and then expands to include their neighborhoods.  While this procedure does not necessarily describe the structure of the entire network, it can yield representative samples of the network \cite{marsden1990network}. Standard random sampling methods can be used to obtain egocentric network data and generalize the results to a larger population \cite{marsden1990network}. Examples of applications of egocentric sampling procedures include the network items in the General Social Survey \cite{burt1984network} and networks obtained through crawling online social networks \cite{gjoka2011practical}.

Numerous algorithms have been developed to calculate subgraph counts. They can be classified in two categories, those that use exact counting \cite{chen2006nemofinder,kashani2009kavosh,grochow2007network,schreiber2005mavisto,hovcevar2014combinatorial} and those that use sampling \cite{wernicke2006fanmod, omidi2009moda, kashtan2004efficient, bhuiyan2012guise, wang2014efficiently}.
Exact algorithms require knowledge, processing, and storage of the entire graph.  In contrast, we provide estimators for subgraph counts using only samples of network data. It is important to note that our methods {\it extend support for sampling} to all previous exact subgraph enumeration techniques that do not support it by design. This is possible by applying exact counting in each egonet and combining the egonet calculations with our Role Occupancy or Unique counting estimators, as described below.  Because of this property, the methods introduced here can be considered complementary to (rather than competitive with) exact counting methods.

mFinder \cite{kashtan2004efficient} was the first algorithm for the estimation of subgraph count using edge sampling. However, it is computationally intensive and scales poorly with the size of subgraphs (up to 6-node). FANMOD \cite{wernicke2006fanmod} uses a node sampling approach and has improved computational complexity vs mFinder (up to 8 nodes). Compared to our approach, mFinder and FANMOD are limited by the fact that they require a uniform independence sample of edges and nodes (respectively) for unbiased estimation. MODA \cite{omidi2009moda} is another sampling algorithm that uses a pattern growth tree approach. While MODA is not as fast as FANMOD, it is able to find larger motifs \cite{wong2012biological}. However, the method does not provide guarantees about bias. GUISE \cite{bhuiyan2012guise} is an algorithm that uses MCMC to sample graphlets, estimating 3-, 4-, and 5-node connected induced subgraphs. This can be used for constructing a graphlet frequency distribution and is much faster than a counting-based approach. Finally, \cite{wang2014efficiently} developed two algorithms, PSRW and MSS, to estimate induced subgraph counts. PSRW uses appropriately re-weighted 
random walk samples and is shown in simulations to have lower estimation errors than FANMOD and GUISE, the latter due to not rejecting samples. MSS, a generalization of GUISE, jointly estimates induced subgraphs of size $k-1$, $k$, and $k+1$ for $k \geq 4$.   Unlike GUISE, PSRW, and MSS, our methods additionally support annotated attributes, and either induced or non-induced (and maximal or non-maximal) subgraphs. In \Sec{sec:simulation} we show that for the estimation of 3-node directed, and 4-node undirected subgraphs our methods outperform PSRW \cite{wang2014efficiently} by up to an order of magnitude for the same sample size.

In our recent work \cite{gjoka13_CliqueEstimation}, we presented statistically principled estimators that count clique subgraphs of all sizes in a graph using ego-centric sampled network data. In this paper, we extend those estimators to support counting all types of subgraphs contained within an egonet. We also examine the effect of network sampling method on estimation error, compare against the state-of-the-art technique for subgraph counting, and demonstrate an application of our estimators to a rare sample of Facebook egonets collected across the whole social graph.

\section{Definitions and Basic Concepts}
\label{sec:notation}

Let $G=(V,E)$ be a undirected or directed graph, with $N=|V|$ nodes and $|E|$ edges.
We associate with the vertices of $G$ a vector $X$ of discrete states, features, or attributes (referred to generically as covariates), such that $X_i$ is the state of the $i$-th vertex of $G$.  Without loss of generality, we denote the possible states of $X_i$ by the integers $1,2,...,p$; in the case for which $p\rightarrow \infty$, it will be possible for our purposes to limit ourselves to the subset of observed states (of which there can be at most $N$).  In a typical social network context, $X$ will represent a categorical or ordinal covariate such as gender or level of education, or else a Cartesian product of such covariates (e.g., gender by level of education, gender by race, etc.).  It is even permissible for $X$ to indicate positional characteristics (e.g., having concurrent partnerships), so long as these are available via the measurement scheme described below.  In some cases, we may be 
interested in graphs whose vertices lack distinguishing characteristics.  In such instances, we take $X$ to be a vector of $1s$.  We refer to the attributes $X$ of $V$ as \emph{annotations}, and say that a graph is \emph{annotated} when it is associated with some attribute vector $X$; a graph without such an association is said to be \emph{unannotated}.

\begin{figure}
\centering
\includegraphics[width=0.40\textwidth]{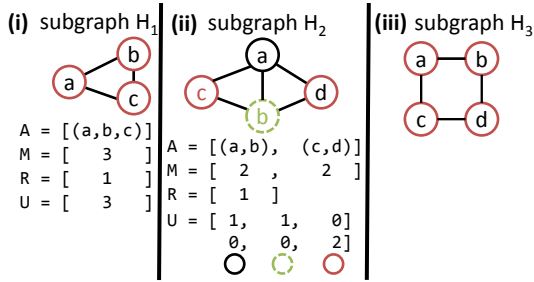}
\caption{Calculation of $A$, $M$, $R$, $U$ for supported subgraphs $H_1$, $H_2$. An example of a non-supported subgraph is $H_3$.  }
\label{fig:subgraph}
\end{figure}

\subsection{Objective}

Our goal is to count the number $C_U$ of subgraphs or induced subgraphs within $G$ of a given attribute composition $U$ that are isomorphic to a particular graph; we define this formally as follows.  
Let $H$ be an unannotated graph of order $h$, representing the structure to be counted, with the directedness of $H$ matching that of $G$.  Denote by $A$ the automorphism orbits of $H$, with $M$ the multiplicities of those orbits (such that $M_i$ is the number of positions within orbit $A_i$ of $H$).  For measurement purposes, we will be particularly concerned with orbits whose members are adjacent to the rest of $H$, for which we introduce specific terminology. 
For undirected $H$, we refer to an orbit $A_i$ as being a \emph{spanning orbit} if any vertex $v \in A_i$ is adjacent\footnote{Note that $H$ may contain loops, and we do not require that a vertex be tied to itself to belong to a spanning orbit.} to all $v' \in H \setminus v$.  
In the directed case, we likewise refer to $A_i$ as being a \emph{semi-spanning orbit} if all $v$ in $A_i$ have semi-edges to all $v' \in H \setminus v$; an \emph{out-spanning orbit} if all v in $A_i$ are adjacent to all $v' \in H\setminus v$; and an \emph{in-spanning orbit} if all $v' \in H \setminus v$ are adjacent to all $v \in A_i$.  Obviously, a graph $H$ containing a spanning or semi-spanning orbit has semi-diameter less than or equal to 2 (although its diameter in the directed case may be larger), though it may contain any number of vertices.  It is this family of graphs with which we are concerned.

While $H$ itself determines the structural form of the subgraph to be counted, we are also concerned with the attributes (or composition) of the vertices within it.  Let $a$ be the number of automorphism orbits of $H$.  We then define a \emph{composition matrix}, $U$, for $H$ to be a $a \times p$ matrix of non-negative integers whose row sums are equal to $M$ (and, therefore, whose total sum is equal to $h$), with $U_{ij}$ indicating the number of members of orbit $A_i$ whose vertex states are equal to $j$. \Fig{fig:subgraph}(ii) shows an example for subgraph $H_2$ that consists of three node attributes represented by colors red, green and black. We calculate $A$, $M$, and $U$ as follows. $H_2$ has two automorphism orbits : the first consists of nodes a, b whereas the second consists of nodes c,d. Hence the multiplicity of both automorphism orbits is two. In the composition matrix $U$ we assign one column to each attribute. We then count the number of nodes of a given attribute in each automorphism orbit.

For the unannotated case, we have $p=1$ and $U=M$ (as an $a \times 1$ matrix); likewise, for the case in which $a=1$ (e.g., cliques), $U$ is a single row-vector corresponding to the number of H-members having each $X$ state.  In the trivial case of unannotated $H$ with $a=1$, $U$ is simply equal to $h$, the order of $H$, as seen in the example of \Fig{fig:subgraph}(i). Typically, however, $U$ will be more elaborate, and it provides the general mechanism by which we will indicate the specific instances of $H$ whose prevalence we seek to estimate.

\subsection{Measurement Assumptions}

We presume that our data comes in the form of an \emph{egocentric network sample} $Y'_1,...,Y'_{n'}$, which is a probability sample of $n'$ egonets from $G$; since the data may be sampled with or without replacement, we denote the unique elements of the sample in arbitrary order by $Y_1,\ldots,Y_n$, with $n'\ge n$. We define the complete egocentric network (or egonet) of vertex $v_i$ (ego) to be $v_i$ together with an appropriately chosen neighborhood of $v_i$ in $G$ (alters), as well as all edges among the indicated vertices (both ego and alters).  The neighborhood over which the egonet is defined (denoted $N^e(v_i)$) is a condition of the measurement process, and may vary; we consider four scenarios:
\begin{itemize}
 \item $G$ is undirected, and $N^e(v_i)$ includes all $v' \in V$ such that $v_i$ and $v'$ are adjacent (\ie $N^e(v_i)=Neigh(v_i)$).
 \item $G$ is directed, and $N^e(v_i)$ includes all $v' \in V$ such that ${v_i,v'}$ is not a null dyad (\ie $N^e(v_i)=Neigh^+(v_i) \cup Neigh^-(v_i)$).
 \item $G$ is directed, and $N^e(v_i)$ includes all $v' \in V$ such that $v_i$ is adjacent to $v'$ (\ie $N^e(v_i)=Neigh^+(v_i))$.
 \item $G$ is directed, and $N^e(v_i)$ includes all $v' \in V$ such that $v'$ is adjacent to $v_i$ (\ie $N^e(v_i)=Neigh^-(v_i))$.
\end{itemize}

In addition to the egonet structure, we assume that the relevant vertex attributes (i.e., $X$ values) are also known for both egos and alters.  The egonet of $v_i$, then, can be represented formally by the vertex set, edge set, value tuple
\begin{equation}
Y_i = (V_i, G[V_i], X[V_i]),
\end{equation}
\noindent where $V_i=N^e(v_i) \cup v_i, G[V_i]$ is the subgraph of $G$ induced by $V_i$, and $X[V_i]$ is the subvector of $X$ corresponding to the vertices of $V_i$.  Our sample is represented by a vector $S$ of length $n$ whose elements index the unique sampled egos, and a vector $p$ of length $n$ whose elements index the {\it inclusion probability} for each sampled ego.  Thus, if $S_i=j$, this implies that the $i$th member of the sample is $v_j$, and $v_j$ was included in the sample with probability $p_i$.  We use this notation in the development that follows.

The choice of egonet neighborhood, $N^e$, determines the subgraphs or induced subgraphs that may be counted using our approach.  In the case of undirected graphs, the subgraphs that may be counted are all $H$ containing at least one spanning orbit.  For directed graphs, three possibilities exist: with $N^e(v_i)=Neigh^+(v_i) \cup Neigh^-(v_i)$, we may count any $H$ containing at least one semi-spanning orbit; if $N^e(v_i)=Neigh^+(v_i)$, we may count any $H$ containing at least one out-spanning orbit; and, finally, if $N^e(v_i)=Neigh^-(v_i)$, we may count any $H$ containing at least one in-spanning orbit.  Intuitively, this is because we require at least one vertex within each copy of $H$ that will contain that copy within his or her egonet - otherwise, we cannot measure it.  We note that additional constraints on $Y_i$ beyond those discussed here (e.g., degree constraints) may place additional constraints on measurable $H$; here, we consider the case in which sampled egonets are measured completely and exactly.

\subsection{Sampling methods}
As noted above, we assume that our egonets comprise a probability sample of the egonets in $G$; that is, (i) we can treat each ego as being included in the sample with known probability, and (ii) the probability of sampling any vertex $v$ is positive for all $v\in V$.  Our estimation supports many sampling methods, including the following.

\smallskip\noindent\textbf{Uniform Independence Sampling (UIS)}, where nodes are sampled independently with equal probabilities. 

\smallskip\noindent\textbf{Weighted Independence Sampling (WIS)}, where nodes are sampled independently with probability proportional to a known weight~$w(v)$.

\smallskip\noindent\textbf{Simple Random Walk (RW)}~\cite{Lovasz93} selects the next-hop node~$v$ uniformly at random among the neighbors of the current node~$u$. 
On a connected and aperiodic graph, RW samples node $v$ with a limiting distribution proportional to its degree~$\deg(v)$. 

Sampling may occur with or without replacement; we indicate these distinctions where they affect estimation.  Note also that, in practice, samples drawn using link-trace methods e.g. Metropolis Hastings Random Walk \cite{gjoka2011practical} or Weighted Random Walk \cite{kurant11_magnifying}, may closely approximate UIS or WIS, and may be employed as well.  We provide an example of this approach in \Sec{sec:facebook}.

\subsection{Neighbor Labeling}

When egonet $S_i$ of ego $i$ is sampled, it may or may not be possible to uniquely identify $i$'s neighbors (in the sense of knowing, e.g., whether $v\in Y_i$ also belongs to some $Y_j$).  When such identification is possible, we say that the sample is \emph{labeled}, otherwise denoting it as \emph{unlabeled}. \Fig{fig:labeling} shows the effect of labeling in an example graph in which egos 7, 4, and 6 are sampled from graph $G$. 
If the sample is labeled, we can discern that the three sampled egonets contain only 1 unique copy of subgraph $H$: $\{7,2,4\}$.  In the unlabeled case, however, we know only that egonet $7$ contains one instance of subgraph $H$, and egonet $4$ contains a second instance.  As we will show, estimation is possible in both cases; however, labeled samples provide additional information that can be leveraged to reduce sampling error.

\section{Estimation}
\label{sec:estimation}

Given a choice of $H$ and $U$, we estimate the number of subgraphs or induced subgraphs having the appropriate structure within $G$,$X$ 
from an egocentric network sample $Y'_1,..,Y'_{n'}$ with unique elements $Y_1,\ldots,Y_n$; in the case of sampling with replacement, $n$ may be less than $n'$. We propose two families of approaches: one, based on role occupancies, that does not require unique identification of neighbors; and one, based on unique subgraph counts, that leverages them.

\subsection{Role Occupancy Method}

Given $H$, $A$, and the choice of egocentric neighborhood used in the measurement of the graph, we define a given orbit of $A$ to be \emph{observable} iff membership of some $v_i$ in such an orbit implies that the associated subgraph (isomorphic to $H$) must be contained in $v_i$'s egonet.  As noted above, our method is applicable to any $H$ with at least one such orbit (given choice of $N^e$).  Let $R$ be a vector containing the indices of the observable orbits in $A$, such that $A_{R_i}$ is observable for all $i$; denote the length of this vector by $r$ (the number of observable roles). For example in the subgraph $H_2$ of \Fig{fig:subgraph}(ii) only orbit 1 that contains nodes $a,b$ is an observable orbit.   We then define the (observable) \emph{role occupancy degrees} of a vertex, $v_i$, to be the vector $d_{i U}$ such that $d_{i U_j}$ is the number of $U$-composition subgraphs or induced subgraphs of type $H$ for which $v_i$ occupies role $R_j$.  It is important to note that, given $Y_i$, $d_{i U_j}$ can be determined exactly for all observable roles, as this is the basis of our technique.

Given the above, we may likewise define the role occupancy degree sums,
\begin{equation}
 D_{U_j} = \sum_{i=1}^N d_{i U_j},
\end{equation}
\noindent which aggregate role occupancies across the graph.  Since each copy of $H$ with composition $U$ has exactly $M_{R_i}$ nodes in the $i$th observable role, it follows that $D$ is deterministically related to the count $C_U$ of such graphs.  Let $m_i=M_{R_i}$ be the multiplicity of the $i$th observable orbit.  Then,
\begin{equation}
 \sum_{j=1}^r D_{U_j} = C_U \sum_{j=1}^r m_j,
\end{equation}
\noindent and so
\begin{equation}
 C_U = (\sum_{j=1}^r D_{U_j})/(\sum_{j=1}^r m_j).
\end{equation}

Since the sum of multiplicities is a constant, we can obtain an unbiased estimate of $C_U$ immediately from any unbiased estimators of the corresponding degree sums.\footnote{Note that sums of unbiased estimators are unbiased estimators of the corresponding sum, a property that follows from linearity of expectation.} A natural way to achieve this is via Horvitz-Thompson (H-T) estimation.  Define
\begin{equation}
 \est{D_U} = \sum_{i=1}^n \sum_{j=1}^r d_{S_i U_j} / p_i
\end{equation}
\noindent to be the H-T estimator of $\sum_{j=1}^r D_{U_j}$, where $p_i$ is the node inclusion probability of $i$ (i.e., the probability of $i$ appearing in the sample at least once).  It then follows that
\begin{equation}
 \est{C_U} = \est{D_U} / (\sum_{j=1}^r m_j)
\end{equation}
\noindent is an unbiased estimator of $C_U$.  Variance estimates can be worked out from H-T theory in the usual manner, exploiting the fact that $\Var(\est{C_U}) = \Var(\est{D_U} / (\sum_{j=1}^r m_j)^2)$.

For designs where the probability that any two observed nodes, $j$ and $k$, are both included in the sample is known, unbiased estimators of the subgraph estimator variance are given by the general form \footnote{This is a direct application of Eq. 6 of \cite{thompson:bk:2002}, p54.}
\begin{multline}
\est{Var}(\est{C_U}) = \sum_{j=1}^n \left(\frac{1}{p_j^2}-\frac{1}{p_j}\right)\left(
\frac{ \sum_{i=1}^{r} d_{S_j U_i}  }{ \sum_{i=1}^{r} m_i } \right)^2\\
 + 2 \sum_{j=1}^n \sum_{k=j+1}^n \left(\frac{1}{p_jp_k}-\frac{1}{p_{jk}}\right)\left(
\frac{ \left[\sum_{i=1}^{r} d_{S_j U_i} \right]\left[\sum_{i=1}^{r} d_{S_k U_i}\right]}{ (\sum_{i=1}^{r} m_i)^2}\right),
\end{multline}
$p_{jk}$ above is the probability of both $j$ and $k$ appearing in the sample.  
For designs such that $p_{jk}$ cannot be readily determined, the generalized H-T estimators of form $\eqn{\ref{e_genestvar}}$ below can be employed.

An important special case arises when sample inclusion probabilities are unequal and known only up to a constant factor (i.e., some $w_j \propto p_j$). Given that joint inclusion probabilities are not available here, an adaptation of the Brewer and Hanif (B-H) variance estimator \cite{brewer.hanif:bk:1983} leads to the following general form:
\begin{equation}
\est{Var}(\est{C_U}) = \left(\frac{N-n}{n(n-1)N }\right)\sum_{j=1}^n\left(\frac{n \sum_{i=1}^{r} d_{S_j U_i}/w_j}{\sum_{k=1}^n \sum_{i=1}^{r} m_i/w_k}-\est{C_U}\right)^2 
\label{e_genestvar} 
\end{equation}
The \hbox{B-H} estimator is generally biased upward \cite{thompson:bk:2002}, and is hence a conservative estimate of measurement error, but does not require joint inclusion weights.

\subsection{Unique Counting Method}

When alters can be uniquely identified across egonets, it becomes possible to estimate $C_U$ by counting unique copies of $H$.  Let $c_u$ be the count of unique copies of H with composition U in the sample, such that an ego belongs to an observable role in each copy.  An H-T estimator of $C_U$ is then immediately given by
\begin{equation}
 \est{C_U} = \sum_{i=1}^{c_u} 1/\pi_i
\end{equation}
\noindent where $\pi_i$ is the probability that the $i$th unique copy appears in the sample. 

As an H-T estimator, the counting estimator is unbiased, and it will generally be more efficient than the role occupancy estimator.  It may, however, be much harder to implement, and in particular its space complexity is much greater.  Obviously, it also requires labeling information.  Given these tradeoffs, both estimators are of potential merit in practice.

\subsection{Inclusion Probabilities}

We have provided estimators of subgraphs counts for either labeled or unlabeled egonet samples.  To use them, it remains only to determine the inclusion probabilities of nodes or subgraphs ($p$ and $\pi$, above).  These quantities depend on the sampling design; we here provide examples for some common and important cases for sampling of OSNs in particular but also other arbitrary graphs.

\paragraph{Node inclusion probabilities}

The simplest case for node inclusion probabilities is that in which egos are sampled uniformly at random from the population (UIS).  The inclusion probabilities depend upon the total number of samples drawn ($n'\ge n$), and whether samples are drawn with or without replacement.  In the with-replacement case, an arbitrary node $j$ fails to be selected on any given draw with probability $1-1/N$, and hence is ultimately included with total probability $p_j=1-(1-1/N)^{n'}$.  When sampling is performed without replacement, a total of $n'=n$ of the $N$ available nodes are drawn, any of which could be $j$.  The resulting inclusion probability is thus simply $p_j=n'/N$.

When the probability of inclusion on any given draw is unequal, total inclusion probabilities may depend on the details of the sampling mechanism.  In the common case of independent with-replacement sampling with unequal probabilities (WIS), the probability of including node $j$ can be determined from the probability of obtaining $j$ on any \emph{given} draw, $p'_j$, by $p_j=1-(1-p'_j)^{n'}$.  Without-replacement inclusion probabilities with unequal are not easily summarized, but computational tools such as \cite{tille.matei:sw:2012} can be employed to obtain them.

In some cases the per-draw inclusion probability may be unequal and known only up to a constant factor (i.e., $w'_j \propto p'_j$).  This situation is common in e.g. random walk sampling of OSNs, where vertices are often sampled (approximately) independently with replacement, proportional to degree.  In such cases, approximating $p'_j$ by the Hansen-Hurwitz \cite{hansen1943theory} estimator $p'_j \approx w'_j \left(\sum_{k=1}^{n'} 1/w'_k\right)/(n'N)$ (where the sum is over all observations, including repetitions) is a practical alternative.

\paragraph{Subgraph inclusion probabilities}

The probability of sampling a subgraph is equal to the probability that at least one member of at least one observable role in the copy is selected as an ego.  The total number of such opportunities is $\sum_{j=1}^r m_j$; the resulting inclusion, however, may depend on which egos occupy the roles in question.  In the case of uniform sampling with replacement (with total draws $n'$), the inclusion probability is
\begin{equation}
\pi_i = 1 - (1-(\sum_{j=1}^r m_j)/N)^{n'},
\end{equation}
\noindent while the corresponding without-replacement probability is
\begin{equation}
\pi_i = 1 - \prod_{k=0}^{n'} (N-k-\sum_{j=1}^r m_j)/(N-k).
\end{equation}

When nodes are drawn non-uniformly, the situation can be more complex, but one case is fairly straightforward.  Let $p'_j$ be the per-draw sampling probability for the $j$th member of copy $i$, under non-uniform independence sampling of nodes.  Then the total inclusion probability for the $i$th copy of $H$ is
\begin{equation}
\pi_i = 1- (1-\sum_{j=1}^{\sum_{k=1}^{r} m_k} p'_j)^{n'}.
\end{equation}

\subsection{Implementation considerations}
\label{subsec:implementation}
The estimation of the absolute subgraph count requires the graph size $N$ and the enumeration of subgraphs for each sampled egonet. In the cases when $N$ is not known a priori, \cite{Katzir2011} provides estimators that work with sampled network data. In general, the exact enumeration of subgraphs is a hard problem. In our approach we avoid enumeration over the whole graph, enumerating only within each sampled egonet. It is important to note that (1) each separate computation can be accelerated because our estimators require subgraph counts only for each ego net; this changes the typical complexity of subgraph counting from $O(f(N))$ to $O(f(\mathrm{deg}_m))$ where $\mathrm{deg}_{m}$ is the maximum degree (often constant in $N$).
 Likewise, (2) the computation on each egonet is independent and thus can be parallelized. Additionally, we can always use the fastest state-of-the-art enumeration tool for the given subgraph type, which speeds up the estimation process. Further, our approach can build on top of existing motif enumeration tools that do not support sampling. For example, in the simulation section we use the  maximal clique listing method by \cite{eppstein2011listing}, the 4-node and 5-node graphlet count method Orca by \cite{hovcevar2014combinatorial}, and a customized subgraph search count by Sage \cite{Sagestein2008} for the 3-node directed subgraphs. Neither of these tools supported subgraph counting with sampling and all of them employ the fastest known algorithms to enumerate the subgraphs that they support.

The space complexity of the Role Occupancy method is $O(n)$, where $n$ is the number of unique egonets sampled. On the other hand, the Unique Counting method requires $O(C_U)$ which can be quite large depending on the subgraph type and the graph structure. To give an example, there are $\sim$627 trillion distinct 5-node ``star'' subgraphs in the network ``web-Google'' and storing them is impractical in this case.

\section{Performance Evaluation via Simulated Sampling}
\label{sec:simulation}

In this section, we evaluate the performance of the estimators with labeled and unlabeled neighborhoods (i.e., unique counting versus role occupancy) via simulated sampling from real-world datasets.  Our results shed light on the relative advantages of these estimators for counting the frequency of a subgraph using sampling.

\begin{table}[t]
\small
\centering
{
\begin{tabular}{|@{}r@{}|@{}r@{}|@{}r@{}|@{}r@{}|@{}r@{}|@{}r@{}|@{}r@{}|}
\hline
    Dataset          & $|V|$   & $|E|$ & \multicolumn{1}{@{}c@{}|}{|E|} &  \multicolumn{1}{@{}c@{}|}{Max.}   &  \multicolumn{1}{@{}c@{}|}{Maxim.} \\
                     &         &  undirect.    & \multicolumn{1}{@{}c@{}|}{directed} &  \multicolumn{1}{@{}c@{}|}{Degree} &  \multicolumn{1}{@{}c@{}|}{Cliq. Sz.} \\
\hline
ca-CondMat~\cite{WWW_SNAP_Graph_Library}         &     21\,362    &     91\,282  &  -            & 279     &  26\\
FB:New Orl.~\cite{Viswanath2009}                 &     63\,392    &    816\,884  &  -            & 1\,098  &   30\\
soc-Slashdot~\cite{WWW_SNAP_Graph_Library}       &     77\,360    &     469\,179 &  828\,161     & 2\,539  &  26\\
soc-sign-Epin~\cite{WWW_SNAP_Graph_Library}      &  119\,129      &  704\,265    & 833\,390      & 3558    &   94\\
email-EuAll~\cite{WWW_SNAP_Graph_Library}        &     224\,832   &   340\,795   &  394\,400     & 7\,636  &   16\\
amazon0601~\cite{WWW_SNAP_Graph_Library}         &     403\,364   &  2\,443\,311 &  3\,387\,224  & 2\,752  &  11\\
web-Google~\cite{WWW_SNAP_Graph_Library}         &  855\,802      & 4\,291\,350  & 5\,066\,842   & 6\,332  &  44 \\
roadnet~\cite{WWW_SNAP_Graph_Library}            &  1\,087\,561   &  1\,541\,512 & -             & 9       &  4\\
youtube-links\cite{Mislove2007}                  & 1\,134\,890    &  2\,987\,623 & 4\,942\,035   & 28\,754 &  17\\
\hline
FB:UCSD~\cite{Traud2011}                        &      14\,948   &     443\,215  & - & 2\,165  &   43\\
FB:UVA~\cite{Traud2011}                         &       17\,196  & 789\,308      & -  & 3\,182  &  42\\
\hline
\end{tabular}}
\caption{Empirical topologies used in Sec.~\ref{sec:simulation} %
}
\label{tab:Topologies}
\end{table}

\subsection{Datasets}
\Tab{tab:Topologies} lists the empirical networks that we use in our evaluation study. It includes several online social networks, an email communication graph, a co-authorship network, a transportation topology and a web graph. Some of the networks have directed edges whereas others have only undirected edges. The former networks are treated as both directed and undirected, depending on the directedness of the subgraph that we are interested in counting $C_U$. For each network we keep the largest connected component\footnote{When the graph is treated as directed we keep the largest weakly connected component.} and we list the number of nodes, number of undirected and directed edges, and the maximum degree and maximum clique size in the undirected graph. The numbers of nodes in the graphs range from tens of thousands to millions.

The list of networks consists of two groups. The first group contains no attributes whereas the second group contains several node attributes. We have selected the node attribute ``gender'' to estimate $C_U$.

\subsection{Error Metrics}
We measure the error of an estimator $\est{C_U}$ with respect to its real value $C_U$ over $k$ simulation iterations using the Normalized Mean Absolute  Error (NRMSE) as follows: \\* $NRMSE(\est{C_U},C_U) = \frac{ \sqrt{ 1/k \cdot \sum_{i=1}^{k}(\est{C_U}-C_U)^2 } }{C_U}$.
In some cases we may want to estimate the count of more than one subgraph e.g. all measurable 5-node undirected subgraphs (see \Fig{fig:5undir}). We summarize the error over all estimations using the Normalized Mean Absolute  Error (NMAE), defined as:
$NMAE(\est{\vec{x}},\vec{x}) = \frac{ \sum(|\est{x}_i-x_i|)  }{\sum |x_i|}$
where~$\vec{x}$ and $\est{\vec{x}}$ are the vectors that correspond to the real and estimated values. $NMAE$ returns the absolute estimation error relative to the true value, averaged over every point in the vector. 

\subsection{Results}

 \begin{figure}[t]
 \centering
 \includegraphics[width=0.49\textwidth]{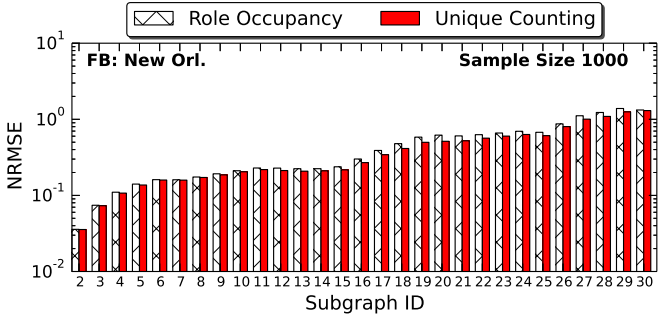}
 \caption{(FB: New Orleans) Subgraph ID $i$ corresponds to the maximal clique of size $i$. NRMSE calculated over 1000 simulated iterations. 1000 egonets are sampled uniformly without replacement.  }
\vspace{-5pt}
 \label{fig:neworleans_Ci}
 \end{figure}

\subsubsection{Role Occupancy (RO) vs Unique Counting (UC)}

We will first compare the performance of our proposed sampling methods, Role Occupancy and Unique Counting. The subgraphs of interest in this comparison will be all order-$i$ maximal cliques.\footnote{A clique is a complete induced subgraph.  A clique that contains $i$ vertices is called an order-$i$ clique. A clique is said to be \emph{maximal} if no higher order clique contains it.}

\Fig{fig:neworleans_Ci} shows the NRMSE for the estimation of all cliques for the topology ``FB: New Orleans'' (from size 2 up to the maximum clique size 30) using a uniform egocentric sample of size $n=1000$. The first observation is that NRMSE is higher for larger clique sizes, probably because those are encountered less often. The second observation is that, as expected, Unique Counting is slightly better than Role Occupancy although the difference is very small for this sample size.

\begin{figure}[t]
 \centering
 \includegraphics[width=0.49\textwidth]{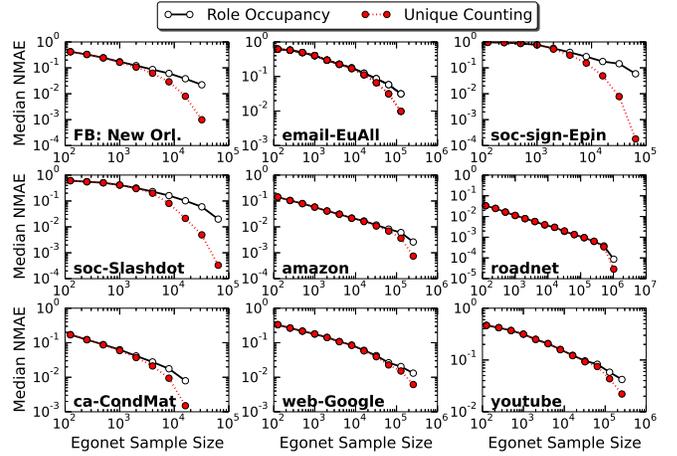}
 \caption{Clique order distribution for real-world topologies. Median NMAE for the estimation of all order-$i$ maximal cliques calculated over 1000 simulated iterations, as a function of the sample size $n$ (uniform sampling without replacement).  }
 \label{fig:allCi_nmae}
 \end{figure}

To get a better understanding of comparative performance, in \Fig{fig:allCi_nmae} we plot the median NMAE of the RO and UC estimators for all order-$i$ maximal cliques on various real-world topologies as a function of sample size. 
We vary $n$ from $125$ to the total size of each graph, allowing us to observe the effects of saturation on measurement error.  We note that for smaller sample sizes, the RO and UC estimators perform equally well.  Beyond a threshold sample size, however, the UC begins to substantially outperform the RO estimator (reflecting the additional information associated with vertex labels). We use \Tab{tab:results} to better interpret these results and shed some light on the causes of the ``threshold'' behavior. \Tab{tab:results} contains for each topology and egonet sample size the average \% of all nodes and \% of all edges  when both egos and neighbors are included. We observe that the UC ``breakaway'' threshold varies for different graphs even when taking into account the total \% of nodes and edges . As an example, the threshold for the network ''soc-sign-Epinions`` is at $n\approx$4\,000, corresponding to $\approx$ 18.1\% of all nodes being  and 34.8\% of all edges being contained in 
some egocentric network sample on average (over 1000 simulations). On the other hand, the threshold for the network ``amazon0601'' is at $n=$64\,000, at which point 80.9\% of all nodes and 63.3\% of all edges have been captured by some egocentric sample on overage.  While saturation aids the UC estimator relative to the RO estimator, the degree of saturation required varies markedly.

\begin{figure}[t]
\centering
\includegraphics[width=0.40\textwidth]{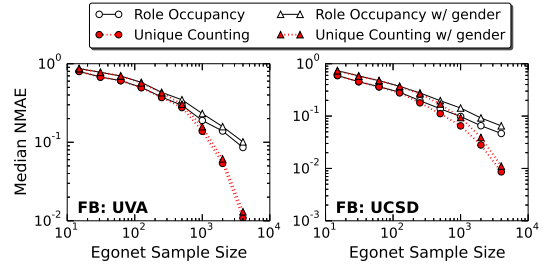}
\caption{Clique order and composition distributions for real-world topologies. Median NMAE for the estimation of all order-$i$ maximal cliques distinguished by gender calculated over 1000 simulated iterations, as a function of $n$ (uniform sampling without replacement). }
\vspace{-5pt}
\label{fig:allCi_Cu_nmae}
\end{figure}

Next, we examine how the addition of node attributes affects the estimation of counts for subgraphs annotated with attributes. For that reason, we simulate the estimation of all order-$i$ maximal cliques that are distinguished by the ``gender`` attribute
\footnote{For example, we estimate all 4 types of order-3 cliques: with 3, 2, 1, or 0 males and corresponding females.}
in Facebook networks UCSD and UVA (see \Fig{fig:allCi_Cu_nmae}). Due to the size and density of these topologies, the egonet sample size is set between $15-4\,000$. \Tab{tab:results} shows the values for the mean \% of nodes and \% of edges  for these egonet sample sizes. As expected from the larger number of values (and smaller counts), estimation of the clique composition distinguished by gender is at least as hard as the estimation without gender. Depending on the composition of the attributes, the estimation w/gender ranges from being indistinguishable  (see FB:UVA) or slightly worse than w/out gender (see FB:UCSD). 

Our results show clear returns to the use of labeled neighborhoods where possible: the UC estimators perform on average as well or better than the RO estimators in all cases.  However, to count the distinct subgraphs the UC estimator needs additional space as discussed in \Sec{subsec:implementation}.
Depending on the topology and the subgraph of interest, the amount of space required to implement the UC estimator might be considerably high. For example, the estimation of all order-$i$ maximal cliques with labeling for the Facebook '09 data samples in \Sec{sec:facebook} requires space that is at least in the order of hundreds of GBytes. Additionally, in some settings (e.g., due to privacy or data collection limitations, particularly offline) is not possible to obtain information on neighbors' identities.  In all these cases, our simulations suggest that the RO estimator can still provide excellent performance, even for very large graphs.

\begin{table}
 \scriptsize
  \centering
  {
\begin{tabular}{|@{}l@{}|@{}r@{}|@{}r@{}|@{}r@{}|@{}r@{}|@{}r@{}|@{}r@{}|@{}r@{}|@{}r@{}|}

\hline
     Dataset \space                    & \%  sampled     &            \multicolumn{7}{c|}{Egonet Sample Size}  \\  \cline{3-9} 

                               &                   &  \xspace \xspace  \xspace  \textbf{250}   & \xspace \xspace \xspace \textbf{500}    &  \xspace \xspace \xspace  \xspace \textbf{1K}   &  \xspace \xspace \xspace \xspace \textbf{2K}      & \xspace \xspace  \xspace \xspace \textbf{4K}     & \xspace \xspace \xspace \xspace \textbf{8K}     &  \xspace \xspace \xspace  \textbf{16K}   \\ 
\hline

\multirow{2}{*}{\begin{minipage}{0.10in}FB:New Orl.\end{minipage}}     
&   \% nodes  & 9.11  &  16.16  &  26.61  &  40.04  &  55.25  &  70.56  &  84.53  \\ 
&  \% edges  & 5.53  &  10.55  &  19.08  &  32.12  &  49.33  &  67.82  &  84.27  \\ 

\hline \hline

\multirow{2}{*}{\begin{minipage}{0.1in}email-EuAll\end{minipage}}     
&   \% nodes  & 0.38  &  0.76  &  1.37  &  2.53  &  4.65  &  8.75  &  16.35  \\ 
&  \% edges  & 0.47  &  0.97  &  1.86  &  3.58  &  6.70  &  12.25  &  21.61  \\ 

\hline \hline

\multirow{2}{*}{\begin{minipage}{0.50in}soc-sign-epinions\end{minipage}}     
&   \% nodes  & 2.23  &  4.05  &  6.86  &  11.31  &  18.14  &  28.61  &  43.77  \\ 
&  \% edges  & 4.31  &  8.36  &  14.21  &  23.30  &  34.84  &  49.10  &  64.32  \\ 

\hline \hline

\multirow{2}{*}{\begin{minipage}{0.1in}soc-slashdot\end{minipage}}        
&   \% nodes  & 3.60  &  6.51  &  11.15  &  18.31  &  28.71  &  43.14  &  61.75  \\ 
&  \% edges  & 1.68  &  3.31  &  6.20  &  11.38  &  19.90  &  33.39  &  53.15  \\ 

\hline \hline

\multirow{2}{*}{\begin{minipage}{0.1in}amazon\end{minipage}}      
&   \% nodes  & 0.80  &  1.60  &  3.15  &  6.15  &  11.72  &  21.59  &  37.48  \\ 
&  \% edges  & 0.42  &  0.85  &  1.69  &  3.35  &  6.56  &  12.60  &  23.37  \\ 

\hline \hline

\multirow{2}{*}{\begin{minipage}{0.1in}web-google\end{minipage}}     
&   \% nodes  & 0.32  &  0.62  &  1.23  &  2.39  &  4.61  &  8.69  &  15.85  \\ 
&  \% edges  & 0.33  &  0.65  &  1.31  &  2.57  &  5.02  &  9.61  &  17.79  \\ 

\hline \hline

\multirow{2}{*}{\begin{minipage}{0.2in}youtube links\end{minipage}}         
&   \% nodes  & 0.13  &  0.26  &  0.52  &  0.98  &  1.88  &  3.51  &  6.46  \\ 
&  \% edges  & 0.10  &  0.21  &  0.46  &  0.85  &  1.71  &  3.27  &  6.19  \\ 

\hline 
\hline

    Dataset & \%  sampled           &            \multicolumn{7}{c|}{Egonet Sample Size}  \\
  \cline{3-9} 
            &                       &  \xspace  \xspace  \textbf{15}   & \xspace \xspace \textbf{31}    &  \xspace  \xspace \textbf{62}   &  \xspace \xspace  \textbf{125}      & \xspace  \xspace  \textbf{250}     & \xspace \xspace \textbf{500} & \xspace \xspace  \textbf{1K}   \\

\hline \hline
\multirow{2}{*}{\begin{minipage}{0.1in}FB: UVA\end{minipage}}     
&   \% nodes  & 7.42  &  14.50  &  25.66  &  41.60  &  60.11  &  75.70  &  86.58  \\ 
&  \% edges  & 2.20  &  4.41  &  8.47  &  15.97  &  28.28  &  44.70  &  63.90  \\ 

\hline \hline
\multirow{2}{*}{\begin{minipage}{0.1in}FB: UCSD\end{minipage}}     

&   \% nodes  & 5.73  &  11.12  &  20.30  &  33.69  &  51.16  &  67.93  &  81.37  \\ 
&  \% edges  & 1.92  &  3.93  &  7.77  &  14.21  &  25.91  &  42.10  &  61.34  \\ 
\hline 
\end{tabular}}
  \caption{
 Uniform sampling without replacement. Total \textbf{\% Nodes Sampled} in the graph when including all egos and neighbors. Total \textbf{\% Edges Sampled} in the graph when including all edges between egos and neighbors. 
}
  \label{tab:results}
\vspace{-5pt}
\end{table}

\begin{figure}[t]
\centering
\subfloat[3-node directed subgraphs]{
\includegraphics[width=0.45\textwidth]{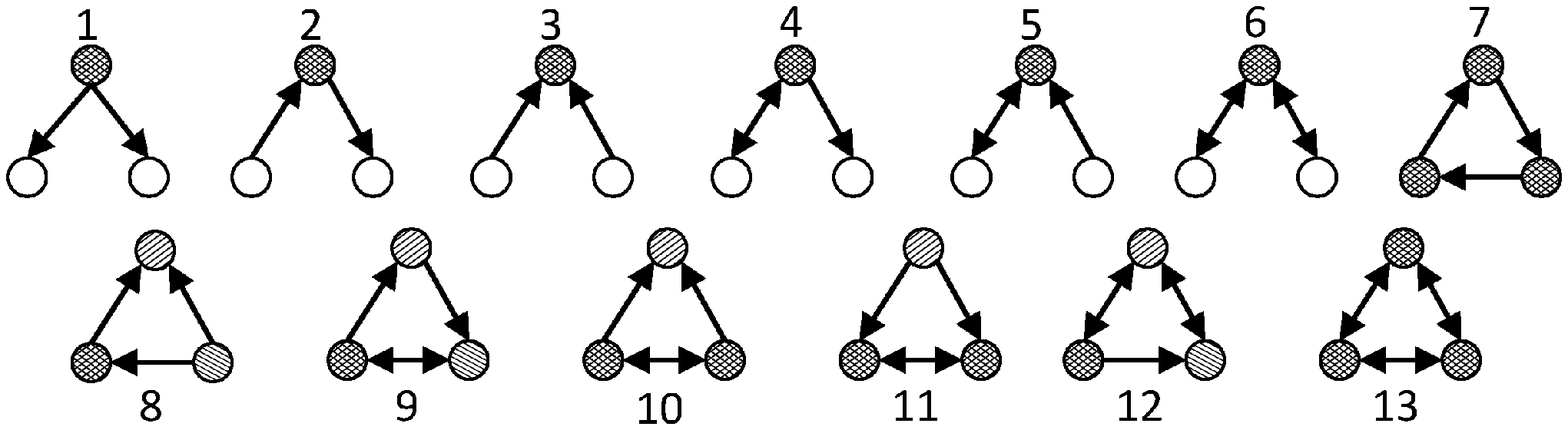}
\label{fig:3dir}
}
\vspace{-5pt}
\vfill
\subfloat[4-node undirected subgraphs]{
\includegraphics[width=0.35\textwidth]{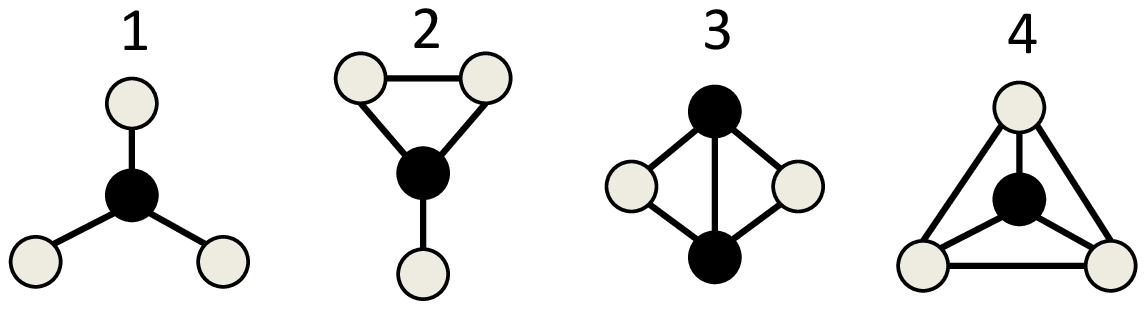}
\label{fig:4undir}
}
\vspace{-5pt}
\vfill
\subfloat[5-node undirected subgraphs]{
\includegraphics[width=0.45\textwidth]{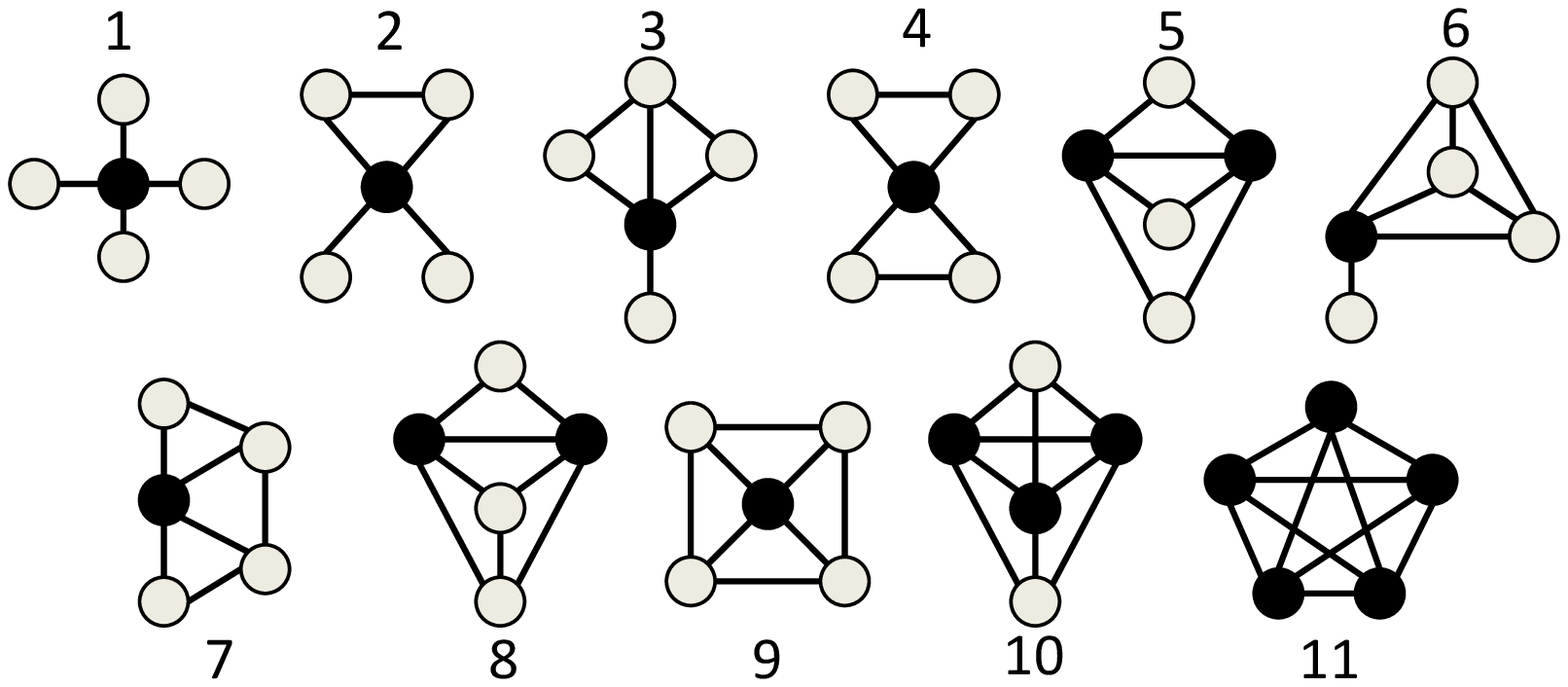}
\label{fig:5undir}
}
\caption{Subgraphs used during simulations. For 4 and 5-node subgraphs, the nodes in black correspond to observable orbits. 3-node subgraphs have multiple observable semi-spanning orbits and the corresponding nodes are colored with a different pattern.
}
\vspace{-5pt}
\label{fig:subgraphs}
\end{figure}

\subsubsection{Effect of sampling method}
\label{sec:effectsampling}

\begin{figure}[t]
\centering
\subfloat{
\includegraphics[width=0.45\textwidth]{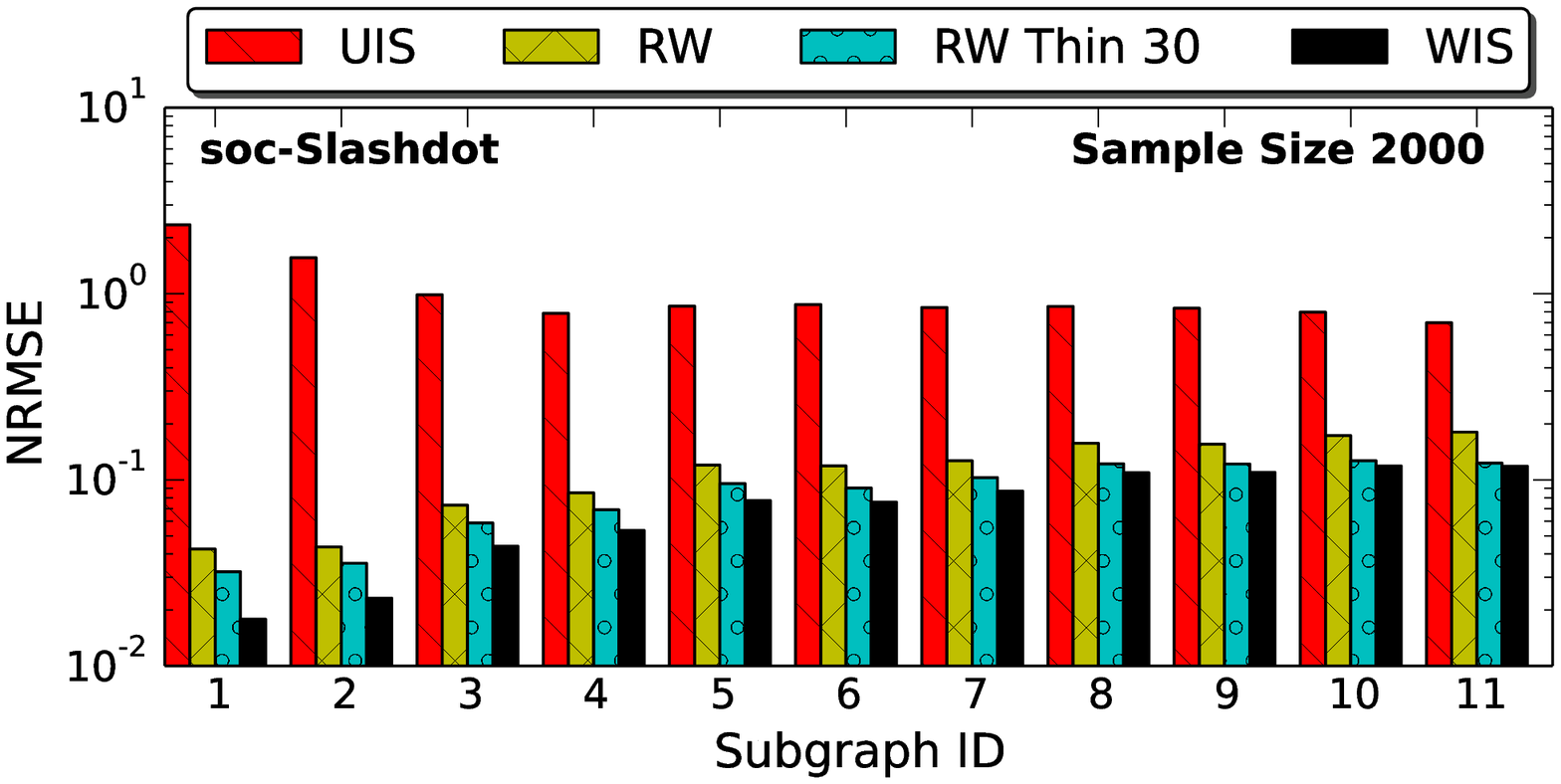}
\label{fig:slashdot_sampling}
}
\vspace{-12pt}
\subfloat{
\includegraphics[width=0.45\textwidth]{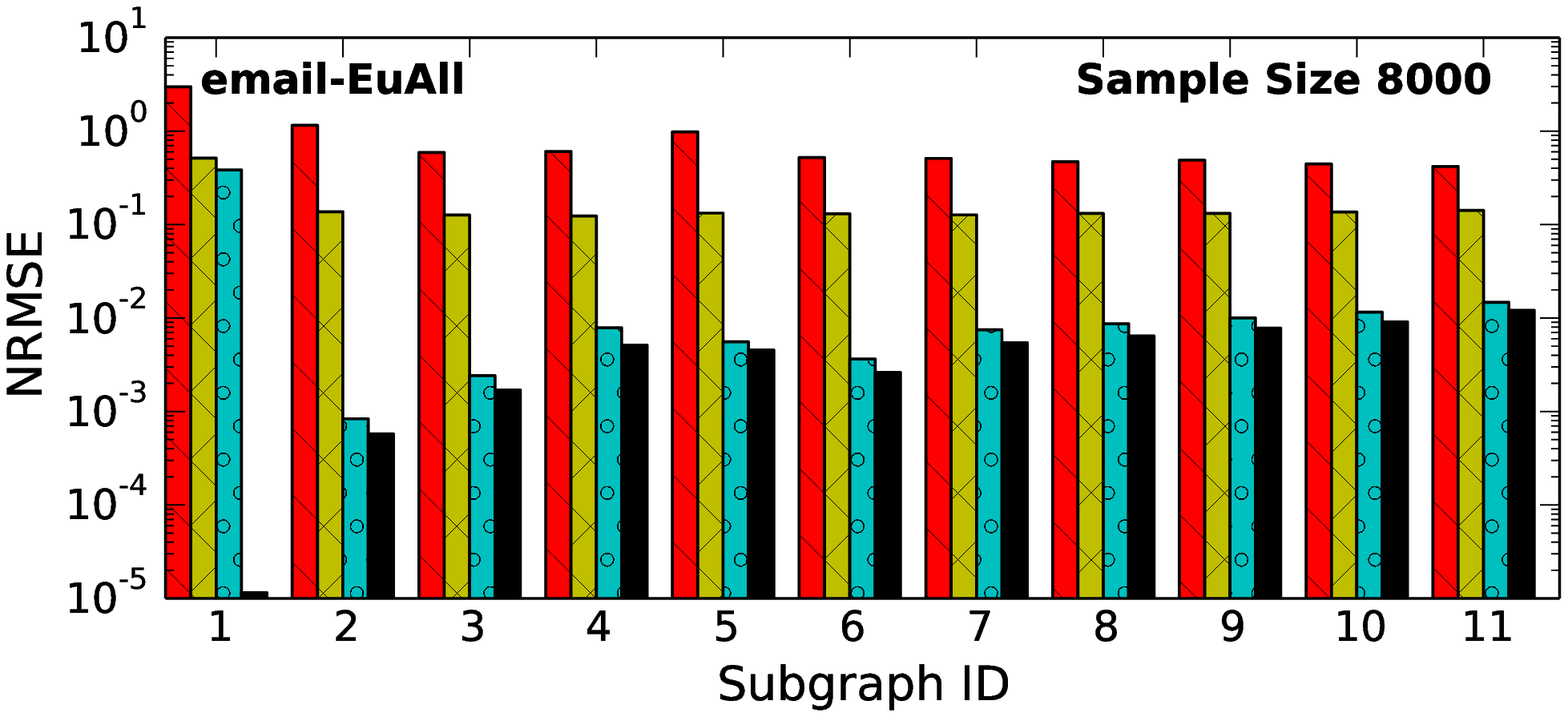}
\label{fig:emailuall_sampling}
}
\vspace{-12pt}
\subfloat{
\includegraphics[width=0.45\textwidth]{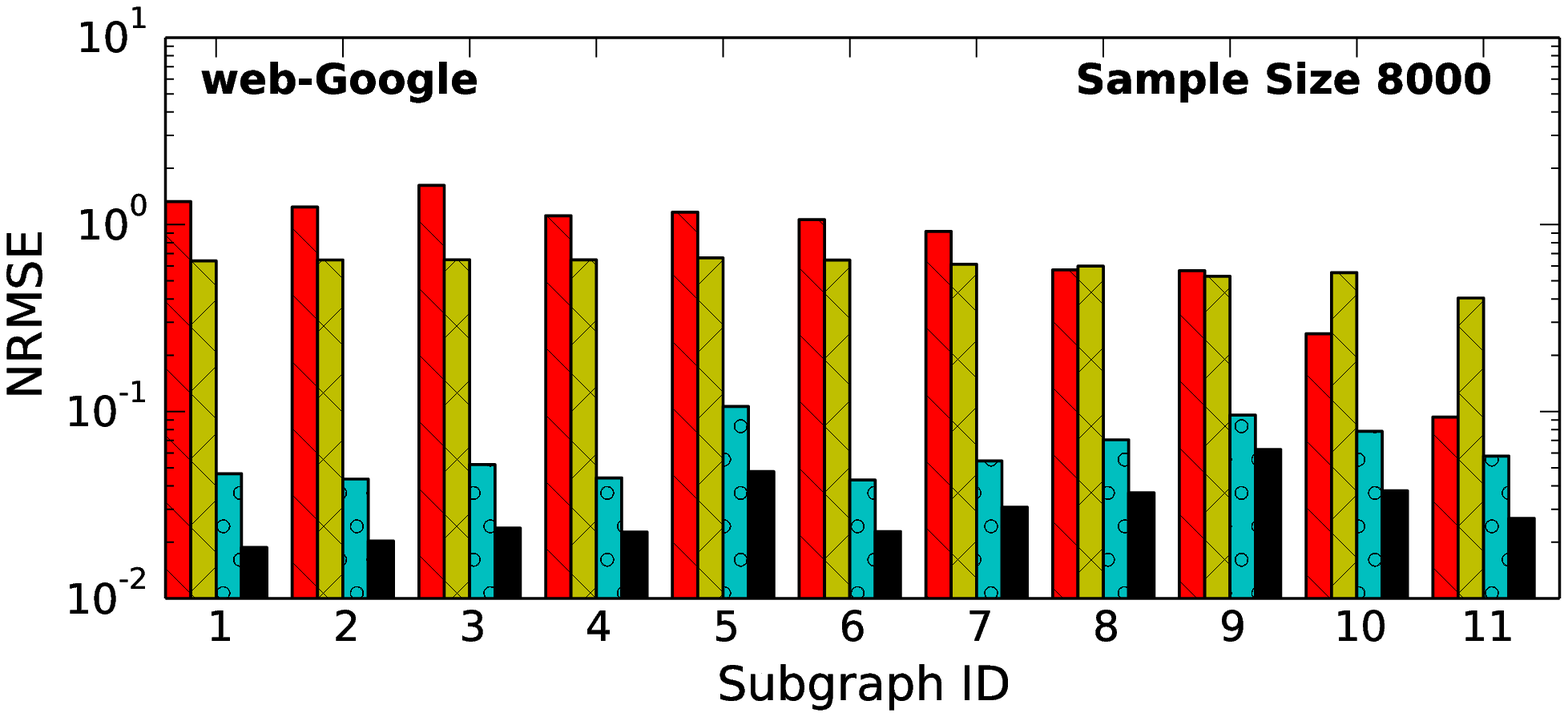}
\label{fig:webgoogle_sampling}
}
\caption{NRMSE for the estimation of all 5-node undirected subgraph IDs (see \Fig{fig:5undir}) in three different graphs averaged over 100 simulation iterations. For each subgraph ID, we list the NRMSE for sampling methods UIS, RW, RW Thin 30, WIS.}
\vspace{-5pt}
\label{fig:nrmse_sampling}
\end{figure}

We now consider the effect of the sampling method on the performance of our estimators. In this evaluation, the subgraphs of interest will be all 5-node undirected subgraphs in \Fig{fig:5undir}. For reasons of spatial complexity we will only use the Role Occupancy estimator. To give an example of the challenges involved with the Unique Counting estimator, there are ~627 and ~10 trillion 5-node subgraphs of type 1 and type 2 correspondingly in the graph ``web-Google.'' The UC estimator requires that we keep track of every distinct subgraph instance which is impractical.

In \Fig{fig:nrmse_sampling} we compare four sampling methods: (1) Uniform Independence Sampling (UIS), (2) Random Walk (RW), (3) Thinned Random Walk, where we collect one egonet every 30 samples in the random walk, and (4) Weighted Independent Sampling (WIS) with the weight of each node set equal to its degree. The NRMSE is shown for the estimation of the subgraph count for all 5-node undirected subgraphs of \Fig{fig:5undir}, in graphs ``soc-Slashdot,'' ``email-EuAll,'' and ``web-Google.'' 

We observe that for UIS, subgraphs with higher number of nodes in observable roles have lower  estimation error. RW samples, however, yield estimation error that is sometimes more than one order of magnitude smaller than UIS in the network soc-Slashdot. The intuition behind this observation is that RW samples are biased toward higher degree nodes which contain a proportionally larger number of subgraph counts. Since we appropriately reweight for the bias, we get a much better (unbiased) estimate of subgraph counts compared to UIS. We should note that the performance boost for RW samples depends on the structure of the network. In cases such as web-Google and email-EuAll, where consecutive RW egonet samples are very correlated and mixing in the network is slow, the performance of RW is comparable with that of UIS. In the latter cases, we observe that by applying thinning in the random walk (every 30 samples) we reduce the correlation of consecutive node to such a degree that we can reach the NRMSE of a WIS sample.

\subsubsection{Comparison with PSRW}

\begin{figure*}
\centering
\subfloat[soc-Slashdot]{
\includegraphics[width=0.90\textwidth]{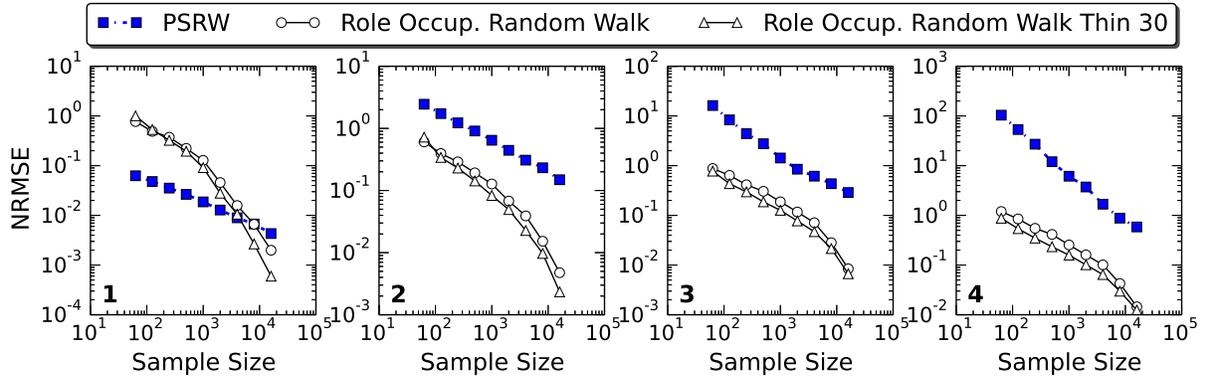}
\label{fig:slashdot_3dir}
}
\vfill
\vspace{-5pt}
\subfloat[amazon]{
\includegraphics[width=0.90\textwidth]{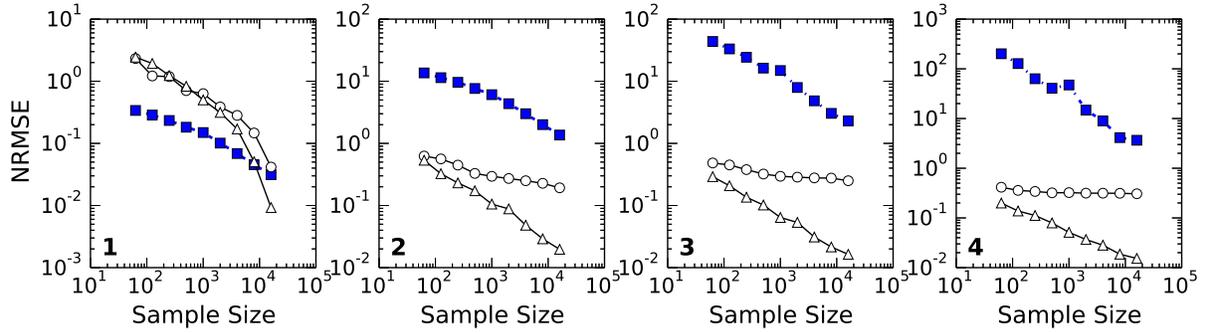}
\label{fig:amazon_4undir}
}
\caption{NRMSE of PSRW and Role Occupancy for the estimation of all 4-node undirected subgraphs supported by our egocentric method}
\label{fig:undir4_results}
\end{figure*}

\begin{figure*}
\centering
\includegraphics[width=0.90\textwidth]{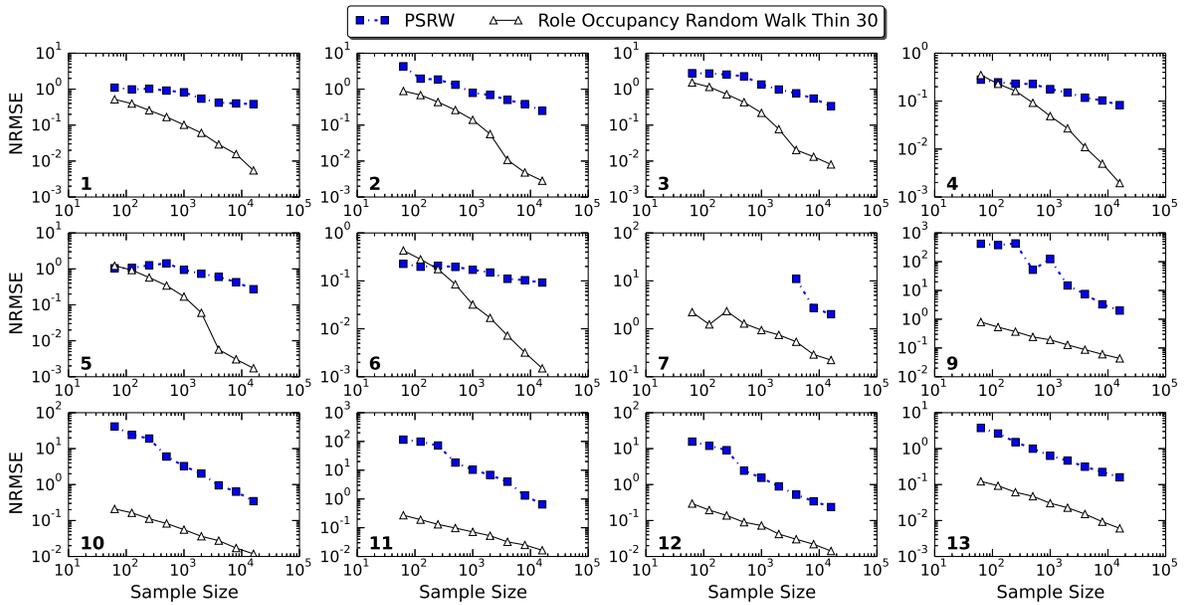}
\caption{NRMSE of PSRW and Role Occupancy (our method) for the estimation of all 3-node directed subgraphs in the network ``youtube-links''}
\vspace{-5pt}
\label{fig:youtube_3dir}
\end{figure*}

To assess relative performance, we compare our Role Occupancy estimator with the state-of-the-art method PSRW \cite{wang2014efficiently}, which was shown to significantly outperform FANMOD and GUISE. We received from the authors of  \cite{wang2014efficiently} source code that implements SRW and PSRW for all 3-node directed subgraphs and 4-node undirected subgraphs.\footnote{The authors of \cite{wang2014efficiently} claim that their method becomes impractically slow for subgraphs of larger size.} 

\Fig{fig:undir4_results} shows the error of the two estimators when estimating all 4-node undirected subgraphs of \Fig{fig:4undir} for networks ``soc-Slashdot'' and ``Amazon.'' In both networks, RO outperforms PSRW by at least an order of magnitude for subgraph IDs 2, 3, and 4. On the other hand for subgraph ID 1, RO  underperforms until sample size $n=10,000$. We also observe that whereas in ``soc-Slashot'' network thinning does not yield any performance improvements, in the ``Amazon'' network thinning the Random Walk every 30 samples yields a significant improvement.

\Fig{fig:youtube_3dir} shows the error of the two estimators  when estimating all \footnote{NRMSE for subgraph type 8 is not shown in \Fig{fig:youtube_3dir} due to similarity with type 9 and economy of space.} 
3-node undirected subgraphs (see \Fig{fig:3dir}) for the network ``youtube-links.'' We observe that the RO method always outperforms PSRW by a significant margin.  RO requires approximately 500-1000 egonet samples to reach NRMSE at or below $10^{-1}$ for all subgraph IDs. On the other hand, in cases of subgraph IDs 7, 8, and 9, the NRMSE for PSRW is as high as $5-10$ after $16K$ samples.

\section{Application to Facebook}
\label{sec:facebook}

\begin{figure}[t]
\centering
\subfloat{
\includegraphics[width=0.40\textwidth]{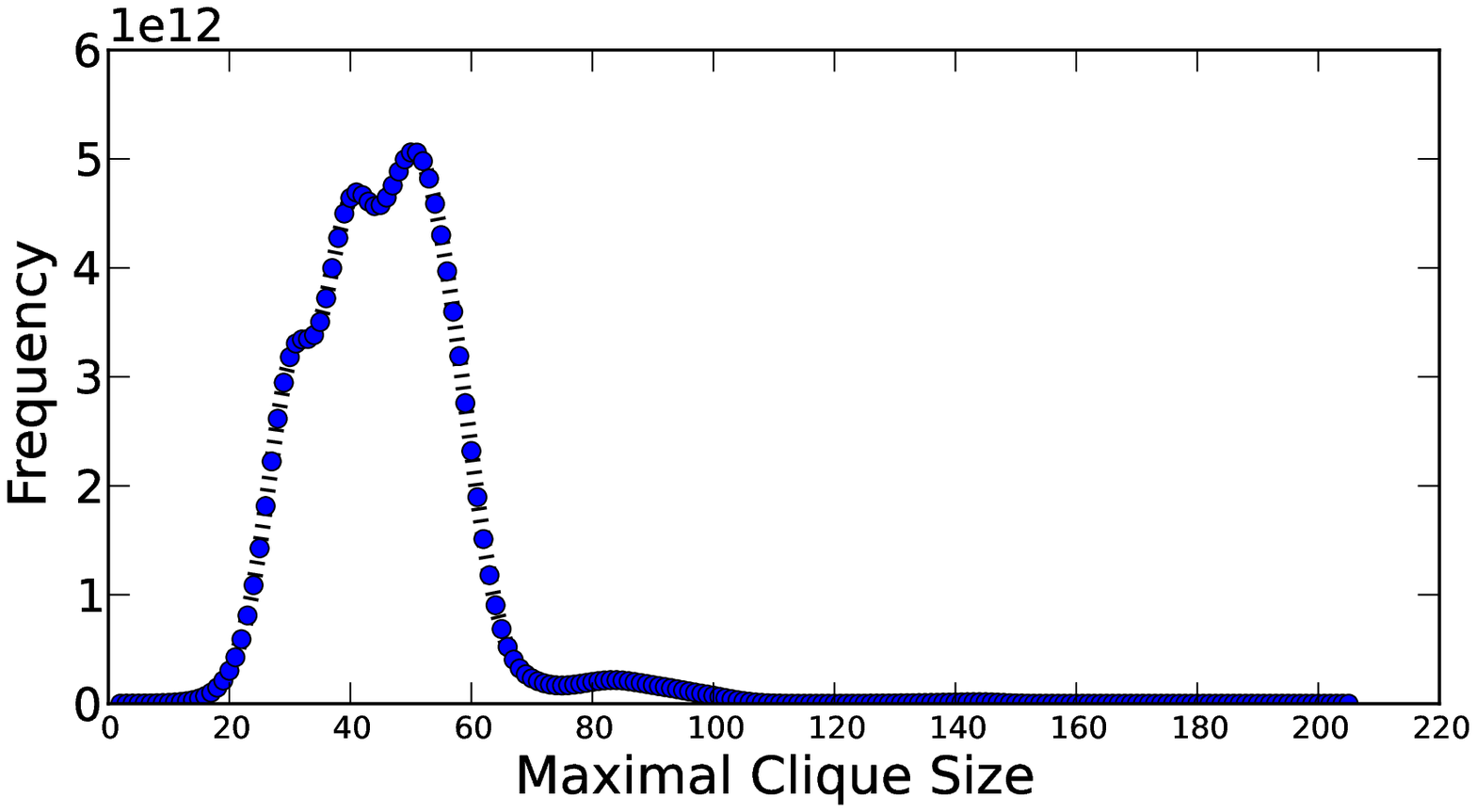}
}
\vspace{-10pt}
\vfill
\subfloat{
\includegraphics[width=0.40\textwidth]{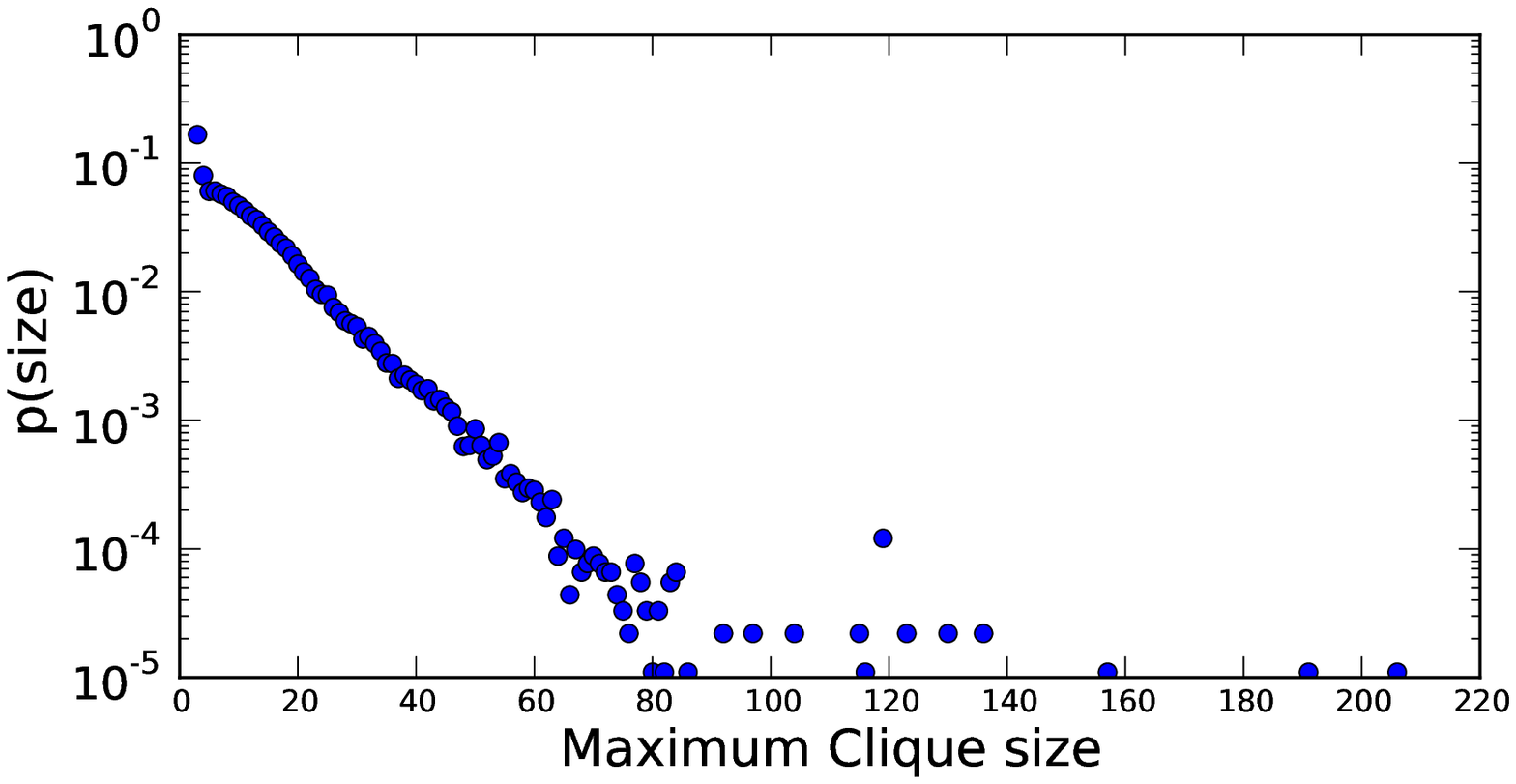}
}1
\caption{Estimated clique size distribution (Facebook social graph '09); top panel shows Role Occupancy estimates of maximal clique frequency, bottom panel shows the distribution of maximum clique sizes by ego.}
\label{fig:fbpublic_Ci}
\end{figure}

\subsection{Dataset Description}
In previous work \cite{gjoka10_walkingfb}, we collected a representative sample of $\approx 1$ million unique Facebook (FB) users by crawling the social graph using a Metropolis Hasting Random Walk (MHRW) method. Subsequently we collected the egonets for $36,628$ unique nodes that were randomly selected from the MHRW sample. This sub-sampling eliminates the correlation of consecutive nodes in the same crawl, similarly to the ``Thin 30'' sample in \Sec{sec:effectsampling}.  The representativeness of this data has been validated against true random samples from the Facebook taken during the same period \cite{gjoka10_walkingfb,gjoka2011practical}.  This sample closely approximates a uniform, with replacement sample of egonets from the publicly visible FB graph. 
In this sample all neighborhoods are uniquely labeled which allows for estimation using either the role occupancy or unique counting estimators. Due to reasons of space complexity, here we use the role occupancy estimators.
We use the population size $N=240M$ which was estimated for this dataset by \cite{Katzir2011} and agrees with the  statistics reported by Facebook during the collection of the dataset (April 2009).

\begin{figure*}[t]
\centering
\includegraphics[width=0.90\textwidth]{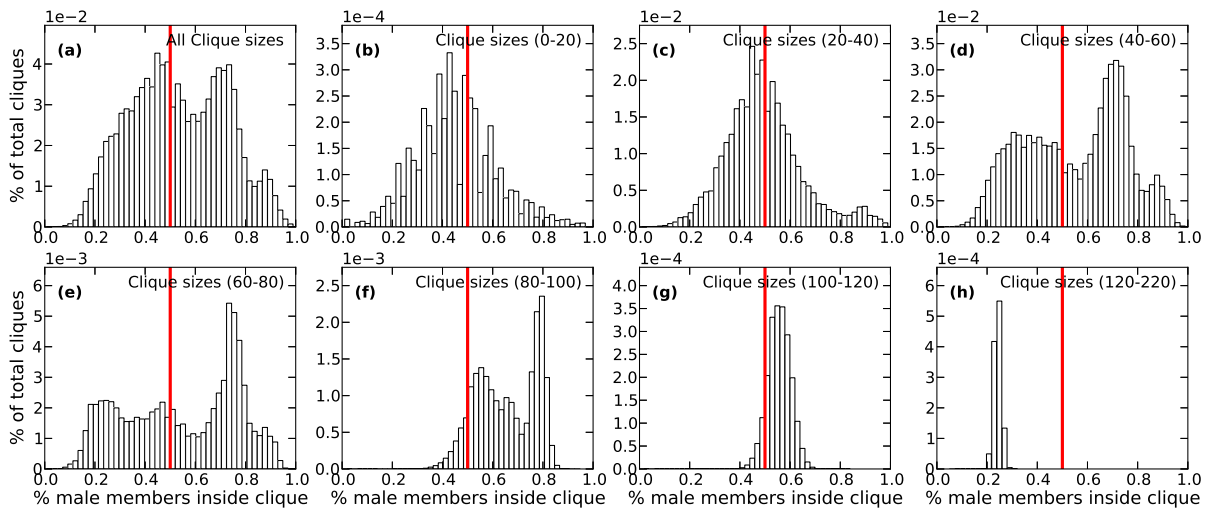}
\caption{Role Occupancy estimates for gender composition of maximal cliques, by order. (FB social graph '09)}
\label{fig:fbpublic_Cu_every20}
\end{figure*}

We complement this egonet sample with gender attributes for each user. We were able to fetch the publicly declared user-declared gender for 90\% of sampled users by crawling the url at \textit{\url{http://graph.facebook.com/userid}}. Additionally, we classified another 9.5\% by a majority rule that uses the first name of each user and a database of the number of times that first names were assigned to males and females. We first used the list of first names from the US Social Security records. If there was no match we then used the list of first names from the population of Facebook users with declared gender.
Last, we used \cite{genderPredictor} to predict the gender for the remaining 0.5\% users with a Naive Bayes classifier, based on the letter composition of first names.

\subsection{Results}

The top panel of \Fig{fig:fbpublic_Ci} shows the estimated distribution of maximal clique sizes over the entire FB social graph.  The FB graph is known to be highly clustered, and it indeed contains many large cliques: the modal clique size is 50, with the largest observed clique containing over 205 individuals.  Interestingly, the form of the clique distribution is neither monotone nor unimodal; significant peaks occur at 32, 41, and 50, with a minor mode near 84.  This suggests substantial heterogeneity in the mechanisms of clique formation, a point underscored by our findings regarding gender (see below).  

Rather more order is shown in the distribution of maximum clique sizes by ego (i.e., the largest clique to which a given individual belongs).  (\Fig{fig:fbpublic_Ci}, bottom panel.)  This shows a monotone distribution with a stable exponential decay over the range that is well-supported by our data.  Membership in moderate to large cliques is thus quite rare, despite their prevalence in the FB graph.

Beyond size distributions, our estimators allow us to examine how the composition of cliques varies across the FB graph.  \Fig{fig:fbpublic_Cu_every20} shows the estimated gender composition of FB cliques for all cliques (panel a) and cliques of varying order (panels b-h).  The $X$ axis in each panel indicates the fraction of clique members who are male, from 0 (entirely female) to 1 (entirely male); a vertical reference line indicates gender parity.  Our results provide clear evidence for strong heterogeneity in the makeup of FB cliques.  We see several distinct modes with characteristic gender frequencies, that occur over specific size ranges.  These include: a ``small equal clique'' mode of near-parity cliques of size 0--40; a 70--80\% male mode for cliques of size 40-100; a 60-80\% female mode for cliques of order 40--80; a second near-parity mode over the small range of sizes 100--120; and a strongly female dominated mode of very large cliques (sizes $>120$).  Although our data does not allow 
us to establish the mechanisms underlying these modes, we speculate that each is the result of a particular collection of social settings (e.g., fraternities or sororities, family groups, schools, or work organizations) that acts as a focus \cite{feld:ajs:1981} for tie formation.  Systematic variation in the gender composition of these settings then leads to corresponding variation in clique composition.  Interestingly, our findings do not corroborate the predictions of \cite{mayhew.et.al:sf:1995} regarding the relationship of clique size to gender homogeneity based on their analysis of face-to-face groups: while they posit a strongly negative relationship between heterogeneity and group size, we find that the FB graph supports a large fraction of near-parity cliques at even quite large orders.  While it is also true that extremely gender-homogeneous cliques become relatively more prevalent at large orders (versus small ones), the phenomenon appears to be uneven and size-specific.  Neither do we observe the 
power-law decay in group sizes reported by \cite{mayhew.et.al:sf:1995} for naturally occurring groups.  Since these prior results were based on observations of cliques in public, face-to-face settings, this lack of replication does not necessarily call into question the validity of the authors' conclusions in their original context; however, it does underscore that the formation of friendship cliques in OSNs may operate very differently from the sorts of groups examined in previous studies.

\subsection{Dataset Release}
We make available at \cite{egonetsample} an anonymized sample of the Facebook egonets that were used in the analysis of this Section. In particular, the dataset contains  $36,628$ egonets, the gender for all egos and neighbors (a total of $\approx 5,6M$ users), and the maximal clique computation without attributes and by gender for all egonets.

\section{Conclusion}
\label{sec:conclusion}

In this paper, we introduced a family of novel unbiased estimators of subgraph counts based on egocentric network samples. We presented two techniques, one of which exploits labeling of nodes (UC) and one which does not require this information (RO). Both techniques are parallelizable, and suitable for use with large graphs.  We evaluated estimator performance via simulated sampling from real-world graphs, showing that both proposed techniques work well and that UC generally outperforms RO as the sample ``saturates'' the graph. We showed that our techniques match or surpass the state-of-the-art method PSRW for subgraph counting and that they can be used for much larger subgraphs. Finally, we demonstrated an application of our estimators to clique composition in OSNs. We applied our methodology to egocentric samples collected in Facebook, which we complemented with gender information, allowing us to estimate the joint size and composition distribution of FB cliques with respect to gender. 
Our results underscore important differences between online and (previously reported) offline group structure, and provide evidence for strong gender heterogeneity in the makeup of FB cliques.

%

\end{document}